\documentclass[]{raa}            

\usepackage{graphicx,times}             
\usepackage{natbib}
\usepackage{amssymb,amsmath}
\usepackage{upgreek}
\usepackage{nicefrac}
\bibpunct{(}{)}{;}{a}{}{,}

\usepackage[a4paper=true,dvipdfm=true,pagebackref=true]{hyperref}
\hypersetup{colorlinks = true, linkcolor = green, anchorcolor = red, citecolor = blue, filecolor = red, pagecolor = red, urlcolor = red}

\newcommand{\comment}[1]{}
\newcommand{\kepler}{Kepler}
\newcommand{\tess}{TESS}
\newcommand{\corot}{CoRoT}
\newcommand{\numax}{\mbox{$\nu_{\rm max}$}}
\newcommand{\Dnu}{\mbox{$\Delta \nu$}}

\newcommand{\muHz}{\mbox{$\mu$Hz}}

\newcommand{\msun}{\!{M_{\sun}}}
\newcommand{\rsun}{\!{R_{\sun}}}
\newcommand{\lsun}{\!{L_{\sun}}}

\newcommand{\diamonds}{\textsc{Diamonds}}
\newcommand{\bestp}{\textsc{BESTP}}

\begin{document}

   \title{{\bestp} --- An Automated Bayesian Modeling Tool for Asteroseismology
}

   \volnopage{Vol.0 (20xx) No.0, 000--000}      
   \setcounter{page}{1}          

   \author{Chen Jiang
      \inst{1,3}
      \and Laurent Gizon
      \inst{1,2,4}
   }

   \institute{Max-Planck-Institut f\"ur Sonnensystemforschung, Justus-von-Liebig-Weg 3, 37077 G\"ottingen, Germany; {\it jiangc@mps.mpg.de}\\
   		\and
   Institut für Astrophysik, Georg-August-Universit\"at G\"ottingen,  Friedrich-Hund-Platz 1, 37077 G\"ottingen, Germany \\
\and 
   		School of Physics and Astronomy, Sun Yat-Sen University, No. 135, Xingang Xi Road, Guangzhou 510275, P. R. China \\
		\and
Center for Space Science, NYUAD Institute, New York University Abu Dhabi, Abu Dhabi, UAE
   		\\
\vs\no}

\abstract{Asteroseismic observations are crucial to constrain stellar models with precision. Bayesian Estimation of STellar Parameters ({\bestp}) is a tool that utilizes Bayesian statistics and nested sampling Monte Carlo algorithm to search for the stellar models that best match a given set of classical and asteroseismic constraints from observations. The computation and evaluation of models are efficiently performed in an automated and a multi-threaded way. To illustrate the capabilities of \bestp, we estimate fundamental stellar properties for the Sun and the red-giant star HD~222076. In both cases, we find models that are consistent with the observations. We also evaluate the improvement in the precision of stellar parameters when the oscillation frequencies of individual modes are included as constraints, compared to the case when only the the large frequency separation is included. For the solar case, the uncertainties of estimated masses, radii and ages are reduced by 0.7\%, 0.3\% and 8\% respectively. For HD~222076, they are reduced even more noticeably by 2\%, 0.5\% and 4.7\%. We also note an improvement of 10\% for the age of HD~222076 when the Gaia parallax is included as a constraint compared to the case when only the large separation is included as constraint. 
}

   \authorrunning{C. Jiang and L. Gizon}            
   \titlerunning{Automated Bayesian Modelling for Asteroseismology}  

   \maketitle

%
%
\section{Introduction}           
\label{sect:intro}


Thanks to the very high quality space-borne photometric observations, such as \corot\ \citep{baglin06} and \kepler\ \citep{borucki10}, asteroseismology has greatly advanced our knowledge about stellar evolution and their inner structures. The detection of oscillations in solar-like and more evolved red-giant stars has particularly led to breakthroughs such as the discovery of differential core rotation and a means fo distinguishing the hydrogen-shell-burning stars from the helium-core-burning ones \citep[see][for a review]{chaplin13}. 

An important part of almost any study of such asteroseismic data is the use of stellar models and the computation of theoretical oscillation frequencies for theses models. In the last few decades, the techniques of data analysis \citep[e.g.,][]{davies16, lund17a, corsaro14,corsaro15}, and also the strategies of stellar modeling \citep[e.g.,][]{serenelli17,silva17} have greatly developed. In particular, by taking individual oscillation frequencies as constraints \citep{metcalfe10, jiang11, mathur12, rendle19}, asteroseismic modeling has proven to be a robust tool to determine fundamental stellar properties, including stellar distances \citep{silva12,rodrigues14}, radii and masses \citep{casa14,pinson14,sharma16}, and most importantly ages for red giants and clump stars \citep{casa16,anders17,pinson18}. Consequently, this enables us to characterize systematically the properties of the hosts of exoplanets through asteroseismic modeling, which in turn provides an unprecedented level of precision in the parameters of the hosted planets \citep{ballard14,camp15,lundkvist16}. Furthermore, with NASA's near all-sky survey the \textit{Transiting Exoplanet Survey Satellite} (TESS) \textit{Mission} \citep{ricker15} launched in 2018, the synergy between asteroseismology and exoplanet science is set to continue to grow \citep{camp18,camp19,huber19,jiang20}, with the mission predicted to increase the number of known solar-type oscillators to a few million \citep{huber18}. The availability of asteroseismic data for large numbers of solar-like oscillators enables us to investigate large ensemble of stars, which is the key to the research of Galactic archaeology \citep[e.g.][]{miglio13, silva18} that studies the structure and evolution of the Galaxy. Under such circumstances, a tool that can carry out modeling and sampling for a large number of stars, in an automated and efficient way, is urgently needed to provide stellar properties with satisfied precision. 

In this paper, we introduce an objective and automated method that performs Bayesian Estimation of STellar Parameters ({\bestp}) by means of fitting stellar models to the asteroseismic data. The method is applicable to the vast number of asteroseismic data expected to emerge from \tess\ and ESA's upcoming mission PLATO \citep{rauer14}. The rest of the paper is organized as follows. In Section~\ref{sc:method}, the detailed computational strategies of the {\bestp} are presented. This is followed by two experimental validations (Section~\ref{sc:experiments}) with the solar data from the Birmingham Solar Oscillation Network \citep[BiSON;][]{chaplin99} and with HD~222076 data from TESS. The conclusion is  at last presented in Section~\ref{sc:conc}.

\section{Computational Method}
\label{sc:method}

Most of our knowledge about stars is gained from observations of layers near the stellar surface. The properties of the interiors of the stars may be inferred based on our best understanding of the constitutive physics to date. Much additional information results from observations of variable stars, in which oscillating waves travel into the deep area of stellar interior and bring information to the upper layer. In particular, the most dramatic example is the nearest star from us: the Sun, in which millions of modes of acoustic oscillation are observed and identified. These modes sample different regions of the solar interior. 
Since distant stars are not spatially resolved, only their modes with spherical harmonic degrees $\ell \lesssim 3$ have measurable amplitudes and can be detected by  \corot, \kepler\ and \tess. In addition to that, the low degree modes can probe deepest into the stellar interior, collecting physical and structural information from the inner core to the outer photosphere.

The precise determination of stellar properties of field stars, especially giant stars, is hard when using classical approaches, such as matching observed spectroscopic parameters to stellar tracks or isochrones. That is because giant stars with different masses evolving in different stages overlap in the observational plane (e.g. in the color-magnitude diagram) within the typical uncertainties of photometry and spectroscopy observations. As a result, the stellar properties determined from these methods are greatly dominated by the statistical errors \citep{serenelli13}. Now we are facing the challenge of precisely determining the fundamental stellar properties using the asteroseismic data, by matching the oscillation modes with the output of theoretical models. In particular, a complete utilization of the analysis of the observations requires that the numerical uncertainties generated for the estimated stellar properties are adequately within the errors of the observational constraints. Many stellar evolution codes, \citep[e.g. {\textsc ASTEC}, {\textsc Cesam2K}, {\textsc GARSTEC}, {\textsc LPCODE}, {\textsc MESA}, {\textsc MONSTAR}, {\textsc YaPSI} and {\textsc YREC};][]{althaus03,pietrin04,demarque08,jcd08a,morel08,weiss08,paxton11,constan14,spada17}, oscillation codes \citep[e.g. {\textsc ADIPLS}, {\textsc GYRE}, {\textsc JIG}, {\textsc LOSC} and {\textsc Pesnell};][]{pesnell90, guenther94, jcd08b, scuflaire08, town13}, and optimization methods \citep[e.g. {\textsc AIMS}, {\textsc AMP}, {\textsc ASTFIT}, {\textsc BeSSP}, {\textsc BASTA} and {\textsc PARAM};][]{deheuvels11, lebreton14, rodrigues14, rodrigues17, silva15, yildiz16, yildiz19, creevey17, serenelli17, mosumgaard18, tayar18, ong19, rendle19} are freely available to the community to determine stellar properties from observed oscillation frequencies. Comprehensive comparisons between 9 well-known evolution codes have recently been carried out for red-giant models and corresponding theoretical frequencies, in the context of the \textit{Aarhus Red Giants Challenge} \citep{jcd20, silva20a}.

The large volume of asteroseismic data that emerges from the \kepler\ and \tess\ missions requires an automated approach to efficiently explore a wide range of model properties in a parallel way, which is the motivation of the development of \bestp. The cornerstone of our model-optimization procedure is based on a code named {\textsc{Diamonds}} \footnote{The package of {\textsc{Diamonds}} is available at https://diamonds.readthedocs.io} \citep[high-DImensional And multi-MOdal NesteD Sampling,][]{corsaro14} that performs model comparison by means of Bayesian parameter estimation. Bayesian statistics has been largely used in asteroseismic analysis \citep[e.g.][]{benomar09, gruberbauer09, kallinger10, handberg11, corsaro13}. Classical techniques such as grid-based modeling, which uses a large set of precomputed grids of stellar models to search for the best-fit model with maximum likelihood, has reached great achievements with its applicability for many stars \citep[e.g.][]{stello09, basu10, basu12, metcalfe10, gai11}. However, methods using Bayesian are typically less computationally expensive than the grid-based modeling as much fewer models are needed. Since the number of stellar models to be evaluated is very large, implementing a Bayesian method can greatly reduce the computational time as well as power for precise determinations of stellar properties. In this section we introduce the stellar evolution and oscillation codes that are used in \bestp\ and how the Bayesian modeling is carried out.

\subsection{Stellar Models}
\label{sc:models}

\bestp\ adoptes the Aarhus STellar Evolution Code \citep[{\textsc ASTEC};][]{jcd08a} and the Aarhus adiabatic oscillation package \citep[{\textsc ADIPLS};][]{jcd08b} to interface with {\textsc{Diamonds}}. {\textsc ASTEC} and {\textsc ADIPLS} were developed for helioseismic analysis, and used by \cite{jcd96} to generate Model~S which is a well-known reference model for solar inversions. The codes are also widely employed for the analysis of oscillations in solar-type and red-giant stars. The input physics of the current modeling include the latest OPAL opacity tables \citep{igl96}, OPAL equation of state in its 2005 version \citep{rog96} and NACRE reaction rates \citep{ang99}. At low temperatures, opacities are obtained from \cite{ferguson05}. Convection is treated under the assumption of mixing-length theory \citep{boh58}. The modelers have the option of including the effect of overshoot and diffusion in the models. However, rotation is not considered.

The input parameters for {\textsc ASTEC} include the initial mass $M$, the initial heavy-element abundance $Z$, and the mixing-length parameter $\alpha$, which are also the parameters to be estimated through the modeling procedure. The hydrogen abundance $X$ is derived based on a Galactic chemical-evolution model \citep{carigi00, pietrin04} from
\begin{equation}
X = 0.748 + -2.4 Z, 
\label{eq:xzrelation}
\end{equation}
which returns $X = 0.706$ when $Z= 0.0173$ and a helium-to-metal enhancement ratio $\Delta Y / \Delta Z = 1.4$.

{\textsc ASTEC} can evolve models from the zero-age main sequence to the tip of the red-giant branch, before the central helium ignition in a helium flash, with a mass-dependent number of internal time steps. Each model evaluation involves stellar model computation and oscillation analysis of these models. For main-sequence star estimations, the computations of stellar evolution tracks terminate before the start of the red-giant branch; for red-giant stars, the termination point is evaluated according to the average large frequency separation for radial modes, $\langle \Dnu \rangle$, which in most cases is a monotonically decreasing function of age \citep{jcd93}, to ensure that the evolution tracks of the stellar models cover the observational constraints. Instead of calculating the oscillation frequencies for every model along an evolutionary track, we also exploit the age dependence of $\langle \Dnu \rangle$. Once the evolutionary track is finished, we compare the calculated $\langle \Dnu \rangle$ with its observed value $\Dnu_{\rm obs}$ to find the best-match model and then calculate frequencies for models within $\pm 10$\% of $\Dnu_{\rm obs}$ from the best-match model. The typical uncertainty on $\Dnu_{\rm obs}$ from one sector of \tess\ data is $\sim$2\% \citep{silva20b}, well below the adopted half range of 10\%. Model evaluations are performed at each iteration of frequency calculations for the selected models, and the best-fit model is returned for each run. Then the frequency calculations are repeated in a narrower range around the best-fit model with a smaller time step, until the time step reaches the nearest age difference along the evolutionary track.

\subsection{Bayesian Modeling}
\label{sc:bayesian}

\bestp\ is aimed to be an automated and efficient modeling tool using asteroseismic data from missions such as \kepler\ and \tess. It is designed to conduct model evaluation by matching the outputs and the observations for any given star, to search for the best-fit model automatically based on the evaluation of each run. Using the observed constraints and the constitutive physics implemented in the models, \diamonds\ can provide a fairly efficient method of finding the optimal models by means of Bayesian parameter estimation and nested sampling Monte Carlo \citep[\textsc{NSMC},][]{sk04} algorithm. The posterior probability is defined according to Bayes' law as 
\begin{equation}
P(\bm{H} | E) = \frac{\mathcal{L}(E | \bm{H})\ P(\bm{H})}{P(E)} 
\label{eq:bayes}
\end{equation}
where $\bm{H}$ represents for the {\it hypothesis} which is, in our analysis, the free parameter vector consisting of the three adjustable model inputs ($M,\, Z,\, \alpha$), and {\it evidence} $E$ corresponds to the calculated model observables $\mathcal M$, for instance luminosity $L$, radius $R$, surface gravity $\log \, g$, effective temperature $T_{\rm eff}$ and individual oscillation frequencies $\nu$. The prior probability distribution $P(\bm{H})$ expresses our knowledge about the free parameters $\bm{H}$ before modeling. The denominator $P(E)$ is the marginal likelihood or Bayesian evidence and is a normalization factor. Under the assumption that observational constraints are statistically independent and normally distributed, the likelihood function $\mathcal{L}(E | \bm{H})$ is given by
\begin{equation}
\mathcal{L}(E | \bm{H}) = \left( \prod_{i=1}^{n} \frac{1}{\sqrt {2 \uppi} \sigma_i} \right) \times \exp \left( -\frac{\chi^2}{2} \right), 
\label{eq:loglikeli}
\end{equation}
with
\begin{equation}
\chi^2 = \sum_{i=1}^n \frac{({\mathcal O}_i - {\mathcal M}_i)^2}  {\sigma^2_i},
\end{equation}
where $\mathcal O$ and $\sigma$ are the observations and errors. By comparing ${\mathcal O}$ and ${\mathcal M}$ the likelihood is then maximized at each sampling iteration. The logarithm of the posterior probability is
\begin{equation} 
\ln [P(\bm{H} | E)] =  \ln[P(\bm{H})] - \ln[P(E)] + \ln [\mathcal{L}(E | \bm{H} )].
\label{eq:logpost}
\end{equation}
As for the prior probability $P(\bm{H})$, a uniform distribution is adopted, which requires the hyperparameters uniformly distributed between the upper and lower bounds for the hypothetical parameters in $\bm{H}$. The search space of the prior distribution is therefore dfined by the hyperparameters before the modeling initiates. To find the hyperparameters for $M$, a preliminary guess is obtained according to the scaling relation \citep{brown91, kjeldsen95, stello09, kallinger10}:
\begin{equation}
\frac{M}{M_{\sun}} \simeq \left( \frac{\numax}{\numax_{,\sun}}\right)^3 \left( \frac{\Dnu}{\Dnu_{\sun}} \right)^{-4} \left( \frac{T_{\rm eff}}{T_{\rm eff, \sun}}\right)^{3/2},
\label{eq:scaledM}
\end{equation}
with $\numax$ the frequency of maximum oscillation power. The symbol ${\sun}$ refers to the solar values, with $T_{\rm eff, \sun} = 5777\, {\rm K}$, $\Dnu_{\sun} = 134.88\, \muHz$ and $\numax_{,\sun} = 3140\, \muHz$ \citep{kallinger14}. Equation~\eqref{eq:scaledM} and other formulations of the scaling relations have been extensively used to estimate the stellar parameters from the observed oscillation frequencies \citep[see, e.g.,][for a review]{hekker20}. \cite{li21} reported that the scaling relation has an intrinsic scatter of 1.7\% in mass ($\sigma_M$) using $\Dnu$ and $\numax$ elaborately generated by the SYD pipeline \citep{huber09,yu18}. However, the typical uncertainty on the estimated mass with \kepler\ and \corot\ data is 5-7\% \citep{huber12,miglio12}. Therefore, the model mass is searched within a $2\,\sigma_M$ range for main-sequence stars, while for more evolved red giants equation~\eqref{eq:scaledM} is less precise and hence a larger search range may be considered. For the hyperparameters of $Z$, first the metallicity [Fe/H] is converted to the abundance ratio $X/Z$ with
\begin{equation}
\mathrm{[Fe/H]} = \log_{10} \left(\frac{X}{Z}\right) - \log_{10} \left( \frac{X}{Z}\right)_{\sun},
\label{eq:metallicity} 
\end{equation}
where the solar abundance ratio $(X/Z)_{\sun}=0.0245$ is adopted  \citep{grevesse93}, and then  equation~\eqref{eq:xzrelation} is used to obtain the preliminary estimations of $Z$ and its errors $\sigma_Z$. The search space covers a $6 \,\sigma_Z$ range. In the case where [Fe/H] is not available, $Z$ is normally  searched within the range of [$0.01$, $0.03$], which corresponds roughly to a space of [$-0.25$, $0.25$] for [Fe/H]. Lastly, the hyperparameters for $\alpha$ are set to be between $1.0$ and $3.0$ with the calibrated solar value equal to $1.96$ (c.f. Section~\ref{sc:sun}). The adopted search space for the hyperparameters is substantial to return reliable outputs for the analysis of solar-type and red-giant stars. However, this method is also applicable to asteroseismic modeling of other types of stars, using different hyperparameters and more importantly with suitable physical ingredients in the stellar evolution code. 

In short, the main steps for \bestp\ with {\textsc{Diamonds}} are as follows:
\begin{enumerate}
\item read observational constraints;
\item set up of models, likelihood function, hyperparameters to be used in the Bayesian inference;
\item set up of drawing algorithm;
\item configuration and start of the nested sampling;
\item calculation and output of results.
\end{enumerate}

\begin{table}[!t]
\begin{center}
\caption{Physical parameters adopted in the modeling tool.}\label{tb:constants}
\begin{tabular}{lcc}
\hline
\hline
\noalign{\smallskip}
Quantity  & Value & Units  \\
\noalign{\smallskip}
\hline
\noalign{\smallskip}
Solar Mass $\msun$ & $1.9890 \times 10^{33}$ & g \\
Solar radius $\rsun$ & $6.95508 \times 10^{10}$ & cm \\
Solar luminosity $\lsun$ & $3.846 \times 10^{33}$ & $\rm g\, cm^2 \, s^{-3}$ \\
Gravitational constant  $G$ & $6.6723 \times 10^{-8}$ & $\rm cm^3 \, g^{-1} \, s^{-2}$ \\
\noalign{\smallskip}
\hline
\noalign{\smallskip}
\end{tabular}
\end{center}
\tablecomments{0.7\textwidth}{The values of $\msun$ and $G$ are based on the measurement of $G\msun$ \citep{jcd05}.}
\end{table}

The \textsc{NSMC} algorithm was developed to efficiently evaluate the Bayesian evidence for any dimensions and to sample the posterior probability distribution for parameter estimation. The detailed \textsc{NSMC} sampling strategies of {\textsc{Diamonds}} are introduced in \cite{corsaro14}. \bestp\ uses three-digit decimal encoding, so in this case there would be a grid of $10^6$ models to be computed within the ranges specified above for the classical grid-based modeling method. In contrast, each run of our method evolves a population of about 2000 to 3000 model tracks to find the optimal set of parameters and, if needed, we perform multiple independent runs with different hyperparameters, for instance with different $Z$ hyperparameters when [Fe/H] is not observed, to make sure that the optimal models identified from the modeling is truly the global solutions. With roughly dozens of oscillation calculations for each track, our method requires about $10^4$ model evaluations, which is nearly 10000 times faster than a complete model grid. Ideally one week of computing time is needed for a $\Dnu \sim 15 \, \muHz$ red-giant star  on a 24-core machine, and much less time (a few days) for a main-sequence star. We also note that, by using a big grid of pre-computed models that covering large ranges of initial model parameters, in principle our method can be applied to multiple observational data sets without repeatedly computing additional models. In addition to that, \bestp\ has the flexibility to incorporate improved physics in the future. 

\begin{figure}
\resizebox{1.0\hsize}{!}{\includegraphics[angle=90]{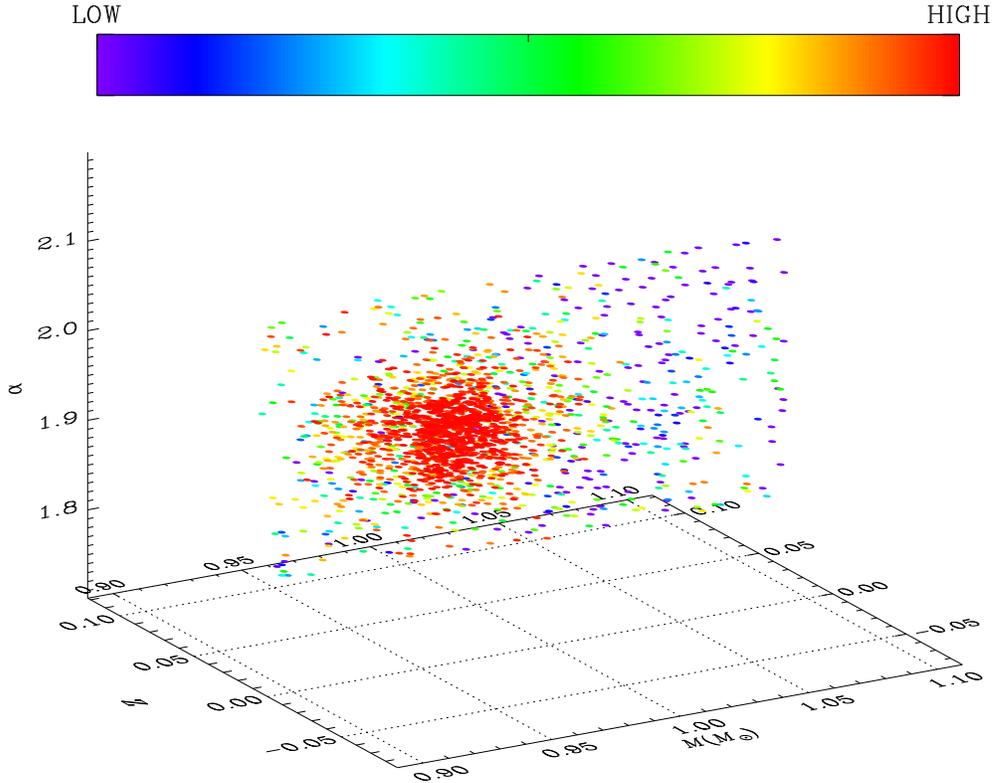}}
\caption{An example of the sampling points drawn in the 3D hyperparameter space of ($M,\, Z,\, \alpha$) by \bestp, for run S2 of the solar modeling experiments, with color-coded logarithmic likelihood values.}
\label{fg:drawS2}
\end{figure}

\section{Model-Fitting Tests}
\label{sc:experiments}

The ultimate goal of \bestp\ is to estimate the optimal stellar parameters by searching for the models that best match the constraints from observations. Including oscillation frequencies in the optimization process adds extra obsrvational constraints to select the best model. In this way, more precise parameters should be yielded compared to the methods that do not use seismic information. In order to test whether \bestp\ returns reliable outputs, in this section we apply it to real observations. First, we begin by applying it to the best observed star, the Sun, for which high-quality disk-integrated data are achievable from multiple observations, such as BiSON, to see whether the tool outputs the known solar properties within acceptable tolerances. Then we apply \bestp\ to a more challenging star, an exoplanet host star that evolves on the red-giant branch, observed by \tess, to validate the efficiency and reliability of the modeling tool.

\begin{figure}
\resizebox{1.0\hsize}{!}{\includegraphics[angle=0]{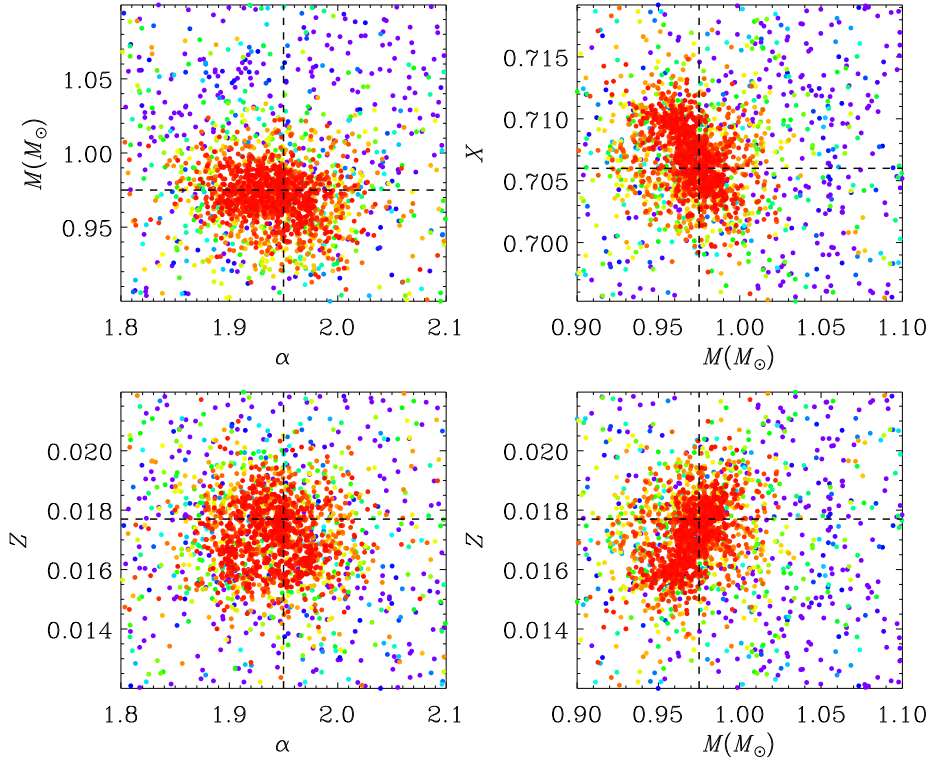}}
\caption{Examples of correlation maps of the the hyperparameters for solar run S2, with the same color-coded logarithmic likelihood values as in Figure~\ref{fg:drawS2}. \textit{Top left}: correlation map of the parameter set ($M, \,\alpha$). \textit{Bottom left}: correlation map of the parameter set ($Z, \, \alpha$). \textit{Bottom right}: correlation map of the parameter set ($Z,\,M$). A correlation can be seen between these two parameters. \textit{Top right}: additional correlation map of the parameter set ($X,\,M$), where $X$ is derived from equation~\ref{eq:xzrelation}. The horizontal and vertical dashed lines represent the locations of the estimated stellar parameters listed in Table~\ref{tb:solar_fit} for the y- and x-axis, respectively.}
\label{fg:mapS2}
\end{figure}

\subsection{Solar Model}
\label{sc:sun}

\bestp\ can only be considered reliable if it results in accurate outputs of the stellar parameters for the star that we study most: our Sun. Here we fit models to real solar data. {\textsc ASTEC} returns stellar properties relative to the solar value. In Table~\ref{tb:constants} we  summarized a set of physical constants adopted in the modeling. The observational constraints consist of basic properties of the Sun: $T_{\rm eff} = 5777 \pm 100 \, \rm K$, $R = 1.00 \pm 0.1 \, \rsun$, $L = 1.00 \pm 0.1 \, \lsun$, $\log g = 4.44 \pm 0.1$, [Fe/H]$ = 0.0 \pm 0.1$, with errors that are comparable to the values expected for stars listed in the \kepler\ Input Catalog \citep{latham05} and the \tess\ Input Catalog \citep{stassun18}. In addition, the seismic quantities $\Dnu_{\sun}$, $\numax_{,\sun}$ and the individual mode frequencies are also used to constrain the model-fitting. The solar oscillation frequencies ($\ell \le 2$)  and their corresponding errors ($\sigma_{\nu} \sim 0.02 \, \muHz$) are obtained from the 22 yr BiSON data set \citep{davies14,hale16}. The results of the modeling are reported in Table~\ref{tb:solar_fit} where estimated stellar properties and their uncertainties are presented for different runs. The 1-$\sigma$ uncertainty correspond to 68.3\% Bayesian credible intervals.

\subsubsection{Individual Frequencies as Constraints}
\label{sc:freCon}

We have performed the modeling without (run S1) and with (run S2) BiSON oscillation frequencies as constraints in two independent runs, while the spectroscopic observations and the asteroseismic parameters $\Dnu$ and $\numax$ are used as constraints in both runs. From Table~\ref{tb:solar_fit}, we see that while the inferred stellar properties are generally consistent between the two runs, using oscillations to constrain the modeling improves the precision of the outputs. This improvement is most significant for the age for which the error decreases by 8\%.

\subsubsection{Search Efficiency}
\label{sc:efficiency}

NSMC initially draws a set of live points (number of live points $N_{\rm live}$ usually ranges from 500 to 2000) from the original prior probability density function (PDF) $P(\bm{H})$. Then at each iteration the model with the lowest likelihood is replaced by a new one, drawn from the prior PDF. This process is repeated until the stopping criterion is met. The set of $N_{\rm live}$ models can be calculated in parallel, as each one of them is independent from the others. However, the  sampling procedure uses information gained from the previous points and hence is not parallelizable. For the solar model experiments, we use  $N_{\rm live}=500$ and  around 2000 sampling points are drawn after the initial set of  live points. An example of the sampling point drawn by \bestp\ is illustrated in Figure~\ref{fg:drawS2}, for run S2 that returns a maximum value of -3.93 for the logarithmic likelihood. The points with high likelihood values are located in the center of the hyperparameter space and hence match best to the observational constraints. Oscillation frequency calculations are certainly computationally demanding, but the number of constraints does not affect the efficiency of the sampling procedure. Figure~\ref{fg:mapS2} presented distributions of the logarithmic likelihood to show relations between the estimated parameters, also for run S2. From the bottom-right diagram of Figure~\ref{fg:mapS2}, we see that there is a correlation between $M$ and $Z$. The reason for this is simple: we use $Z$ to determine $X$ based on equation~\ref{eq:xzrelation}, so that a higher value of $Z$ returns a lower value of $X$ and hence a larger-mass model to match the observations.

The computational efficiency is determined by several aspects, including the computation of the stellar models and oscillation modes, the value of $N_{\rm live}$, the configuration of nested sampling, etc. The bottleneck  is the computation time of stellar models and mode frequencies. The current versions of {\textsc ASTEC} and {\textsc ADIPLS} are not parallelizable. The computation of a solar-type evolutionary track usually takes only a few seconds, but it can take up to $\sim$10 minutes for a red-giant track. The number of sampling points is around 2000 to 3000 for solar-type and red-giant star modeling.

\begin{table}[!t]
\begin{center}
\caption{Results of fits to the solar data.}\label{tb:solar_fit}
\begin{tabular}{lccc}
\hline
\hline
\noalign{\smallskip}
Run  & cal & S1 & S2  \\
\noalign{\smallskip}
\hline
\noalign{\smallskip}
$M$ ($\msun$) & 1.0 & $0.982 \pm 0.019$ & $0.975 \pm 0.012$  \\
$Z$ & 0.0173 & $0.0179 \pm 0.0012$ & $0.0177 \pm 0.0008$   \\
$X$  & 0.719 &   $0.705 \pm 0.003$ & $0.706 \pm  0.002$  \\
$\alpha$ & 1.938 & $1.939 \pm 0.038$ & $1.950 \pm 0.025$ \\
$t$ (Gyr) & 4.57 & $4.92 \pm 0.91$& $5.25 \pm 0.54$ \\
$T_{\rm eff}$ (K) & 5779 & $5782 \pm 28$ & $5782 \pm 27$  \\
$L$ ($\lsun$) & 1.0 & $0.986 \pm 0.030$ & $0.980 \pm 0.023$ \\
$R$ ($\rsun$) & 1.0 & $ 0.992 \pm 0.007$ & $0.990 \pm 0.004$  \\
$\log g$ (cgs) & 4.438 & $ 4.435 \pm 0.004$ & $4.434 \pm 0.002$ \\
$[{\rm Fe}/{\rm H}]$  & $-0.008$ & $0.015 \pm 0.029$ & $0.010 \pm 0.019$ \\
\noalign{\smallskip}
\hline
\noalign{\smallskip}
\end{tabular}
\end{center}
\tablecomments{0.85\textwidth}{BiSON oscillation frequencies are used as constraints in run S2, but not in run S1. Global seismic parameters  $\Dnu$ and $\numax$ are used in both runs. The calibration run `cal' is described in Section~\ref{sc:sun_fit}.}
\end{table}

\begin{figure}
\resizebox{1.0\hsize}{!}{\includegraphics[angle=90]{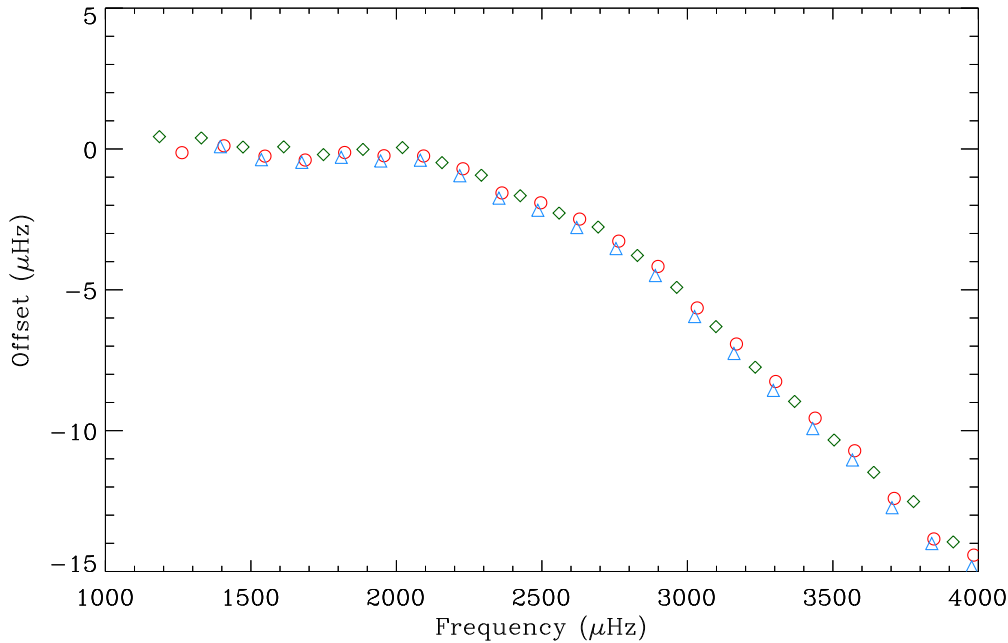}}
\caption{Offset due to the near-surface effects between the BiSON frequencies (circles: $\ell = 0$, diamonds: $\ell =1$, triangles: $\ell = 2$) and the corresponding modes of the best-fit model from run S2 of the solar case. }
\label{fg:surfSun}
\end{figure}

\subsubsection{Surface Effect Correction}
\label{sc:surface}

\begin{figure}
\centering
\resizebox{0.7\hsize}{!}{\includegraphics[angle=90]{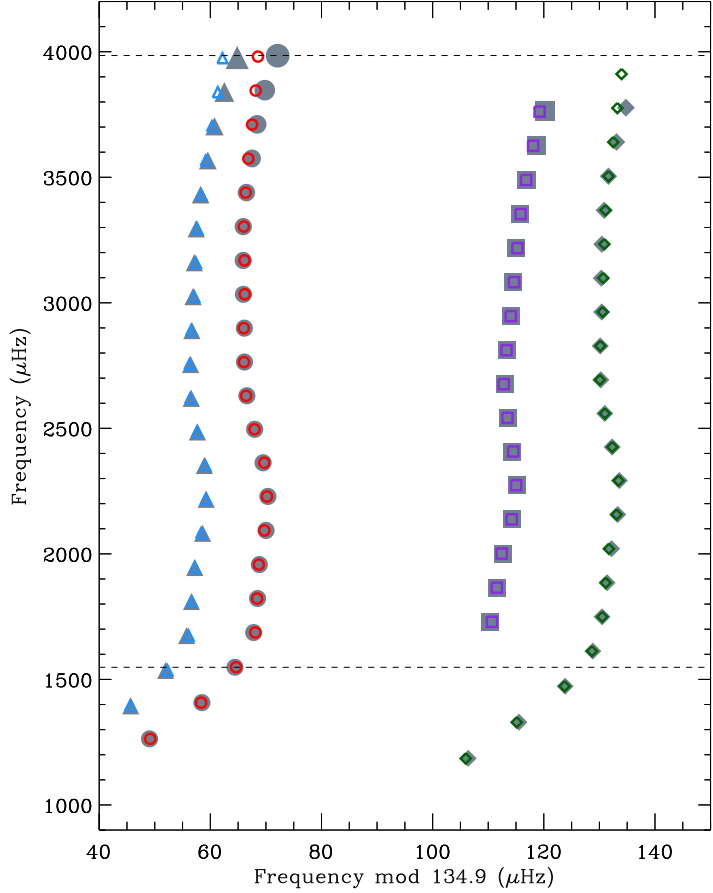}}
\caption{\'Echelle diagram illustrating observed mode frequencies from BiSON data (filled gray symbols) and the representative best-fit model (open colored symbols) from run S2 of the solar case, for radial (circles), dipole (diamonds), quadrupole (triangles), and octupole (squares) modes. Sizes of the BiSON modes are scaled according to the 1-$\sigma$ observation errors. Theoretical modes are corrected for the near-surface effect by utilizing the combined terms defined by Equation (4) in \cite{ball14}. Note that only $\ell < 3$ modes with frequencies inside the dashed lines are used as constraints for the fit, but the best-fit model also matches the $\ell = 3$ modes and those outside the dashed lines.}
\label{fg:echelle_sun}
\end{figure}

The observed low-frequency modes of solar oscillation are from BiSON and are compared with the theoretical frequencies computed by the {\textsc ADIPLS}. It has been established by several studies \citep{jcd88,jcd96,dziembowski88,jcd97} that there is a systematic deviation between the theoretical and observed frequencies due to the deficient treatment of the layers near the surface in the models, known as the \textit{surface effect}. The effect yields a systematic offset of several $\muHz$ between the observed and calculated frequencies (see an example of Figure~\ref{fg:surfSun}) and grows larger with higher frequencies. The offset is universal to modes with different $\ell$ and displays no dependence of $\ell$ (also see Figure~\ref{fg:surfSun}). \cite{kjeldsen08} proposed a method of empirical correction of the surface effects that uses a solar calibrated power law. This empirical method is computationally cheap compared to the three-dimensional radiation hydrodynamics simulations of convection which is theoretically able to reduce the frequency discrepancies caused by the surface effect. The last few years have seen some new approaches emerged to correct the offset \citep[e.g.][]{roxburgh03, gruberbauer12, ball14, sonoi15,houdek17}. \cite{ball17} compared several methods and found that the combined terms of surface correction proposed by \cite{ball14} are slightly superior to other corrections, including the solar-calibrated power law formulated by \cite{kjeldsen08}, when modeling subgiants and early red-giant stars with their individual mode frequencies. Therefore, we adopt the combined correction from \cite{ball14}. An example of the surface corrected frequencies from BiSON data is shown in Figure~\ref{fg:echelle_sun}. Although tiny discrepancies between the corrected frequencies and BiSON data are seen towards high frequency, the mean value of the discrepancy is $0.36$\,\muHz\ corresponding to a mean relative difference of $0.1$\%. We note that only $\ell < 3$ modes are used to constrain the models, but the best-fit model also matches the observations for $\ell = 3$ modes in Figure~\ref{fg:echelle_sun}.

\subsubsection{Discussion}
\label{sc:sun_fit}

In Table~\ref{tb:solar_fit} the estimated outputs for $M$, $R$, $\log g$ and $t$ are given for both runs of S1 and S2, constraining the corresponding properties to $\sim$1.9\%, $\sim$0.7\%, $\sim$0.09\% and $\sim$18\% for S1, and to $\sim$1.2\%, $\sim$0.4\%, $\sim$0.05\% and $\sim$10\% for S2, respectively. This demonstrates that including individual oscillation frequencies as model-fitting constraints indeed increases the precision of the estimated properties, especially for $t$ whose precision is increased by 8\%. However, the estimated $M$, $R$ and $L$ are a little smaller than the known solar values given in Table~\ref{tb:constants}, on account of the underestimated $X$ compared with the calibrated solar model (run 'cal' in Table~\ref{tb:solar_fit}) generated by {\textsc ASTEC}. The calibration is performed by iterating over $\alpha$ and the initial $X$ to get $L_{\sun}$ and $R_{\sun}$ at $4.57$~Gyr, with initial $Z = 0.0173$, which returns $X_{\rm cal}=0.719$ for the calibrated model. Nevertheless, the modeling procedure estimates the values of $X$ by using equation~\ref{eq:xzrelation} which gives $X=0.706$ with $Z = 0.0173$. It takes longer time for the models with smaller initial $X$ to reach the same point on the evolutionary tracks. That is why the estimated ages for the solar model from run S1 and S2 are a little larger than that used in the calibration. The differences in the output properties might in principle be reduced if both $X$ and $Z$ are set free, rather than fixed by the helium-to-metal enrichment ratio given in Section~\ref{sc:models}. However, freeing both $X$ and $Z$ would increase the computation time dramatically, but would not lead to significant improvement. As we can see for the solar cases the discrepancies between the \bestp\ models and the calibrated model are very small ($\lesssim$1\%). Therefore, we conclude that the optimal models returned by \bestp\ satisfactorily approximate the known solar properties.

\begin{table}[!t]
\begin{center}
\caption{Results of model fits for HD~222076.}\label{tb:hd_fit}
\begin{tabular}{lcccc}
\hline
\hline
\noalign{\smallskip}
Run &OBS  & T1 & T2 & T3  \\
\noalign{\smallskip}
\hline
\noalign{\smallskip}
$\Dnu$ ($\muHz$) & $15.6 \pm 0.13$ &$15.6 \pm 0.07$ & $15.9 \pm 0.02$  &$15.9 \pm 0.02$\\
$M$ ($\msun$) &$\cdots$ & $1.154 \pm 0.071$ & $1.174 \pm 0.050$  & $1.163 \pm 0.037$ \\
$Z$ &$\cdots$ & $0.0199 \pm 0.0027$ & $0.0204 \pm 0.0030$   & $0.0201 \pm 0.0027$ \\
$X$  &$\cdots$ & $0.701 \pm  0.007$ & $0.699 \pm 0.007$  &   $0.700 \pm 0.006$ \\
$\alpha$ &$\cdots$ & $1.907 \pm 0.081$ & $1.919 \pm 0.072$  & $1.920 \pm 0.061$\\
$t$ (Gyr) &$\cdots$ & $6.59 \pm 1.46$ & $6.32 \pm 1.10$ & $6.52 \pm 0.82$\\
$T_{\rm eff}$ (K) &$4806 \pm 100$ & $4795 \pm 45$ & $4808 \pm 48$   & $4804 \pm 27$\\
$L$ ($\lsun$) &$9.14 \pm 0.33$ & $9.26 \pm 0.56$  & $9.23 \pm 0.56$ & $9.15 \pm 0.23$\\
$R$ ($\rsun$) &$\cdots$ & $4.415 \pm 0.089$ & $4.394 \pm 0.066$   & $ 4.379 \pm 0.045$\\
$\log g$ (cgs) &$\cdots$ & $3.210 \pm 0.009$ & $3.223 \pm 0.006$  & $ 3.221 \pm 0.005$\\
$[{\rm Fe}/{\rm H}]$ &$0.05 \pm 0.10$ & $0.06 \pm 0.06$ & $0.08 \pm 0.06$  & $0.07 \pm 0.06$\\
\noalign{\smallskip}
\hline
\noalign{\smallskip}
\end{tabular}
\end{center}
\tablecomments{0.85\textwidth}{The observation constraints are listed in the `OBS' column. The \tess\ oscillation frequencies are used as constraints in run T2, but not in run T1. Gaia luminosity, complemented with individual frequencies, are used in run T3. The global seismic parameter  $\Dnu$ is used in all the runs. }
\end{table}

\subsection{HD~222076}
\label{sc:tess}

The Bayesian modeling tool \bestp\ was successfully applied to the red-giant star 46~LMi observed by the Hertzsprung Stellar Observation Network Group (SONG) Telescope \citep{grundahl17} as its first target, to determine the stellar mass and age using asteroseismic modeling \citep{frandsen18}. More recently, as \tess\ released its first asteroseismic data, we analyzed the planet-hosting red-giant star HD~222076 (TIC~325178933) and foretell the destiny of the planetary system by modeling the stellar age and the timing of the engulfment of the planet. In the following sections we present the detailed modeling of HD~222076 with \tess\ asteroseismic data.

\subsubsection{Modeling}
\label{sc:hd_modeling}

\begin{figure}
\resizebox{1.0\hsize}{!}{\includegraphics[angle=0]{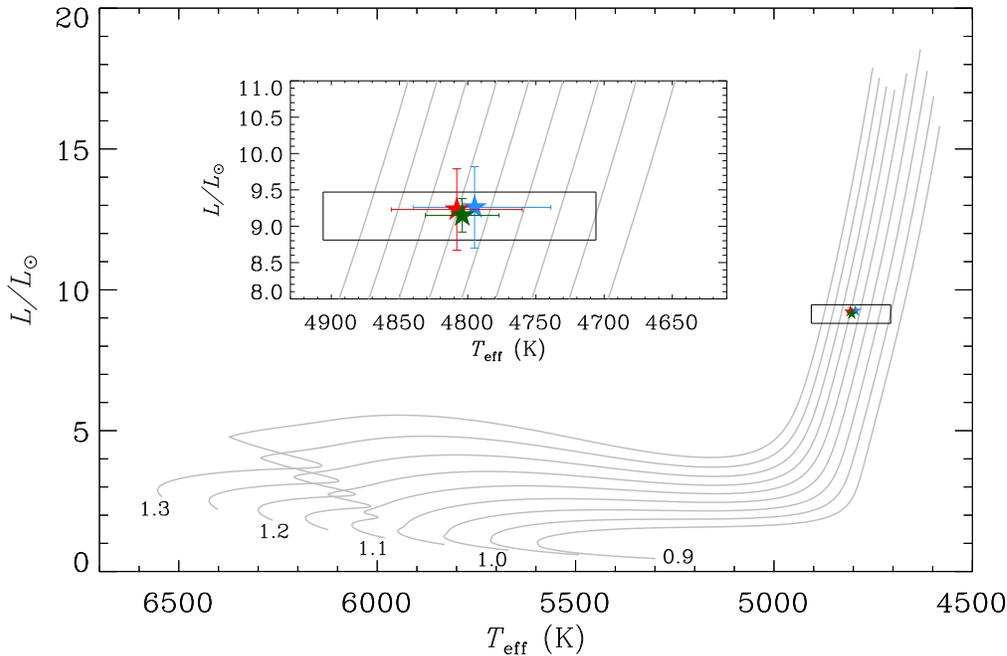}}
\caption{Evolutionary tracks for eight models with different initial masses but same chemical abundance ($X=0.702$ and $Z=0.0191$) and mixing length parameter ($\alpha = 1.917$). The initial masses of the models increase from 0.9 $\msun$ to 1.3 $\msun$ (values presented at the start of every other tracks) with a step of $0.05 \msun$. The small rectangle is the 1-$\sigma$ error box for the observational constraints (i.e. $T_{\rm eff}$ and $L$) of HD~222076. The star symbols indicate the locations of the fitting results (from Table~\ref{tb:hd_fit}) of run T1 (blue), T2 (red), T3 (green). A zoom-in on the area around the observational constraints is also shown in the figure, with 1 $\sigma$ error bars of estimated parameters illustrated.}
\label{fg:tess_tracks}
\end{figure}

The asteroseismic analysis of HD~222076 with \tess\ data was done by \cite{jiang20}. The star is ascending on the red-giant branch with $\Dnu = 15.6 \pm 0.13 \, \muHz$ and $\numax = 203.0 \pm 3.6 \, \muHz$, which leads to a preliminary mass of $1.15 \pm 0.13~\msun$ by adopting a spectroscopic $T_{\rm eff} = 4806 \pm 100 \,$K \citep{wittenmyer16} in equation~\eqref{eq:scaledM}. \cite{jiang20} determined the stellar mass to be 1.12~$\msun$. Therefore, in the modeling process we search for masses in the range between 1.0 and 1.3~$\msun$, rather than using a larger space as described in Section~\ref{sc:bayesian}. The hyperparameters for $Z$ are [$0.01$, $0.03$], covering the observed [Fe/H]$=0.05 \pm 0.10$, and those for $\alpha$ are [1.7, 2.1]. Using a Gaia DR2 parallax of $11.024 \pm 0.022 \, $mas \citep{gaia18} with the bolometric flux $F_{\rm bol} = 3.57 \pm 0.13 \times 10^{-8}$\,erg\,s$^{-1}$\,cm$^{-2}$ obtained from the broadband spectral energy distribution (SED) analysis yields $L = 9.14 \pm 0.33\,L_{\sun}$ (see Section~2.2 of \cite{jiang20} for details of the SED analysis). We have tested \bestp\ for HD~222076 in three independent scenarios with different observation constraints: without (run T1) and with (run T2) individual oscillation frequencies, and with (run T3) both oscillations and $L$. And in all three scenarios the constraints are complemented with the spectroscopic $T_{\rm eff}$ and [Fe/H], as well as asteroseismic $\Dnu$. The HR diagram showing the theoretical evolutionary tracks and the spectroscopic constraints are presented in Figure~\ref{fg:tess_tracks}.

\subsubsection{Results}
\label{sc:hd_results}

The modeling results for the three scenarios (T1, T2 and T3) are listed in Table~\ref{tb:hd_fit}. The results returned by the three runs are generally consistent with each other. However, the uncertainties of the estimated parameters are larger than in the solar case, which is expected for red giants. Masses, radii and ages obtained from T1, in which only $\Dnu$ and spectroscopic observables are used, are with uncertainties at the 6\%, 2\% and 22\% level that are less than half of the values reported by \cite{silva20b} who performed the first asteroseismic investigations for an ensemble of red-giant stars with \tess\, using the same observational constraints as in T1 plus $\numax$. This level of statistic uncertainties is sufficient for asteroseismic study of large ensemble of red-giant stars. Moreover, including individual oscillation frequencies as constraints in the modeling (T2) decreases the uncertainties of the inferred parameters, compared with the run T1, by 2\%, 0.5\% and 4.7\% for $M$, $R$ and $t$, respectively. Further improvement in the precision is seen when the luminosity obtained from Gaia DR2 parallax is included in the fit (T3), as the statistic uncertainties of T3 increase by $\sim$3\%, $\sim$1\% and $\sim$10\% from T1, for $M$, $R$ and $t$, respectively. T2 and T3 returns somewhat higher values of $\Dnu$ than observations, because the estimations of $\Dnu$ are performed before the surface effect correction that can reduce the value of $\Dnu$. Overall, \bestp\ is able to reproduces observations in all the three scenarios. As illustrated in Figure~\ref{fg:tess_tracks}, the estimated stellar properties are located near the center of the error box of the observational constraints, with uncertainties of $T_{\rm eff}$ all below the 1$\sigma$ observation error. However, uncertainties of $L$ from T1 and T2 are higher than the observation error, but it significantly decreases in T3, thanks to the inclusion of Gaia luminosity.

We note that the precisions yielded by T3 are improved by $\sim$8\% for $M$, $\sim$4\% for $R$ and $\sim$24\% for $t$, compared with the consolidated results from different codes and methods by \cite{jiang20} as additional systematic errors are included in their analysis. The computation of models for HD~222076 takes much longer time than the solar case, due to the time-consuming calculations of g modes.

\section{Conclusion}
\label{sc:conc}

This article introduces the automated asteroseismic modeling tool \bestp\ that is based on the {\textsc{Diamonds}} code using Bayesian parameter estimation to perform model comparison. With the successful validations of \bestp\ using real solar data observed by BiSON, as well as the data of a TESS red-giant star HD~222076, we showed the robustness and efficiency of the tool. From the comparisons between observations and models computed by \bestp\, we estimated the fundamental stellar properties, such as mass, radius and age, for the Sun and HD~222076. We validated the significant improvements in precisions of the estimated parameters returned by asteroseismic modeling, when using individual oscillation frequencies as constraints. The uncertainties of estimated masses, radii and ages are reduced by 0.7\%, 0.3\% and 8\% for the solar case and even more noticeably by 2\%, 0.5\% and 4.7\% for the red giant HD~222076.
We also noted an improvement of the estimations by 10\% for the age when the Gaia DR2 parallax is included as constraints compared with the case when only the large frequency spacing is included as constraint.
However, the number of stars for which spectroscopic data as well as parallaxes from Gaia are available is currently much larger than the number of stars with asteroseismic data. An increase in the number of stars characterized by asteroseismology (using \kepler, \tess, later PLATO) will lead to the same increase in  the number of stars with precise stellar parameters.

\bestp\ can be used to determine ages for a large set of stars, which is crucial for various studies, such as  Galactic archaeology and exoplanetary science. Although the computation times for red-giant stars are relatively long, which limits the potential of using asteroseismic modeling for a very large number of evolved stars, 
a smaller sample of asteroseismic targets can still serve as a calibration set for the full set of stars, either using the scaling relations \citep{bellinger19, bellinger20} or deep learning \citep{hon20}. Nevertheless, the computational efficiency could be significantly improved by adopting a pre-calculated model database, thereby reducing  computational time for stellar models and oscillation frequencies. This paper discussed only the random errors. Biases introduced by missing physics etc. is well beyond the present study.

\section{Acknowledgements}

The author gratefully acknowledges the computing time granted by the Yunnan Observatories. This work is funded by the Fundamental Research Funds for the Central Universities (grant: 19lgpy278) and is also supported by a grant from the Max Planck Society to prepare for the scientific exploitation of the PLATO mission.

\label{lastpage}

\end{document}